\documentclass[10pt]{iopart}

\usepackage[english]{babel}

\usepackage{amsfonts} 
\usepackage{amssymb}
\usepackage{cite}
\usepackage{graphicx}

\usepackage{xcolor}
\usepackage{ifthen}
\usepackage{scalerel}

\newcommand{\dsub}[3]{\ensuremath{#1_{#2_{\scaleto{#3}{4pt}}}}}
\newcommand{\eqspacing}{\hspace{15pt}}

\newcommand{\rescale}[2]{\ensuremath{#1 \to #2 \ #1}}

\newcommand{\Nr}[1]{\ensuremath{ \ifthenelse{\equal{#1}{}}{ \mathbf{r} }{ \mathbf{r}_#1 } }}
\newcommand{\Np}[1]{\ensuremath{ \ifthenelse{\equal{#1}{}}{ \mathbf{p} }{ \mathbf{p}_#1 } }}
\newcommand{\Nphi}[1]{\ensuremath{ \ifthenelse{\equal{#1}{}}{ \varphi }{ \varphi_#1  } }}
\newcommand{\Nu}[1]{\ensuremath{ \ifthenelse{\equal{#1}{}}{ \mathbf{\hat{u}} }{ \mathbf{\hat{u}}_#1  } }}
\newcommand{\Nx}[1]{\ensuremath{ \mathbf{x}_{#1} } }

\newcommand{\Nfunctional}{\ensuremath{\mathcal{F}}}

\newcommand{\Npart}[1]{\ensuremath{\frac{\partial #1}{\partial t}}}
\newcommand{\Nint}[3]{\ensuremath{\int_{#1}^{#2} \! \! d#3}}
\newcommand{\norm}[1]{\left|\hspace{-0.25pt}\left| #1 \right|\hspace{-0.25pt}\right|}

\begin{document}

\title[Mean field approach of dynamical pattern formation in underdamped active matter]{Mean field approach of dynamical pattern formation in underdamped active matter with short-ranged alignment and distant anti-alignment interactions}

\author{Dominic Arold and Michael Schmiedeberg}

\address{Institut f\"ur Theoretische Physik I, Friedrich-Alexander-Universit\"at Erlangen-N\"urnberg, Staudtstra\ss e 7, 91058 Erlangen, Germany}

\ead{michael.schmiedeberg@fau.de}

\begin{abstract}
  Many active matter systems, especially on the microscopic scale, are well approximated as overdamped, meaning that any inertial momentum is immediately dissipated by the environment. On the other hand, especially for macroscopic active systems but also for many mesoscopic systems the time scale of inertial motion can become large enough to be relevant for the dynamics. This raises the question how collective dynamics and the resulting states in active matter are influenced by inertia. Therefore, we propose a coarse-grained continuum model for underdamped active matter based on a mean field description for passive systems. Furthermore, we apply the model to a system with interactions that support an alignment on short distances and an anti-alignment on longer length scales as known in the context of pattern formation due to orientational interactions. Our numerical calculations of the under- and overdamped dynamics both predict a structured laning state. However, activity induced convective flows that are only present in the underdamped model destabilize this state when the anti-alignment is weakened, leading to a collective motion state which does not occur in the overdamped limit. A turbulent transition regime between the two states can be characterized by strong density fluctuations and the absence of global ordering.
\end{abstract}

%\keywords{magnetic moment, solar neutrinos, astrophysics}
\noindent{\it Keywords\/}: active matter, underdamped dynamics, collective motion
%\submitto{\JPCM}
%\maketitle
\ioptwocol

\section{Introduction}

Active matter denotes a whole plethora of biological and artificial many-body systems composed of self-propelled, interacting particles \cite{Marchetti2013, bechinger2016active,  doostmohammadi2018active}. Often, microscopic systems are considered in which the constituents move in a viscous environment. Typical examples include protein filaments in motility assays \cite{sumino2012large}, cell colonies \cite{Dombrowski2004, riedel2005self, Rossen2014}, biological or artificial microswimmers \cite{Reinken2018a, Heidenreich2016, Menzel2016}, all surrounded by a viscous background fluid at low Reynolds numbers or binding to an adhesive substrate. These systems are overdamped, meaning the motion of particles is solely governed by the momentary forces acting on them. However, Newton's first law formulates that massive objects resist a change in momentum due to their inertia. While not relevant on the microscopic length scale in solution, inertial effects become relevant in typical macroscopic realizations of active matter like flocks of birds \cite{ballerini2008interaction, ginelli2010relevance} or artificially made massive robots moving on a two-dimensional plate \cite{Kudrolli2008, Bricard2015, Deseigne2012}. For the latter, the influence of inertia on the long time statistics was observed recently for a one particle system \cite{scholz2018inertial}.

There is a wide range of particle based active matter models which explicitly incorporate the microscopic interactions \cite{Menzel2013a, grossmann2014vortex, wensink2012emergent, mccandlish2012spontaneous}. Another approach to such systems is the formulation of continuum models which typically describe systems on a length scale above the particle resolution. Such coarse-grained models are already able to predict experimental observations of microscopic systems like bacterial colonies \cite{wensink2012meso}, ensembles of microswimmers \cite{Heidenreich2016, Ariel2018} or active nematics \cite{chandragiri2019active, mueller2019emergence, doostmohammadi2018active}. However, for other systems where the time scale of inertial motion becomes relevant an extended underdamped description of the dynamics becomes necessary.

In this article we propose a continuum model for classical underdamped active matter by including activity to a dynamical density functional theory for underdamped passive systems \cite{Archer2009}. We use the model to identify distinct states of a system with local alignment and distant anti-alignment interaction of particle orientations and capture the consequences of introducing inertia by comparing the found states to those of the corresponding overdamped limit system. Note that in our approach the actual velocity of a particle that is the cause of inertia effects does not have to coincide with the orientation of a particle that determines the direction of the intrinsic self-propulsion.

The article is structured as follows: In section \ref{sec:model} we derive the model and specify the system. Simulation results on the latter are discussed in section \ref{sec:results} and concluded in section \ref{sec:conclusion}.

\section{Model}
\label{sec:model}
\subsection{Derivation}
In this section we outline the derivation of the general model which follows \cite{Archer2009} concerning the calculation steps and coincides with it when active components are neglected. These additional active terms do not change the principal calculation but give additional contributions in each step. The inclusion of activity follows similar approaches used for overdamped active systems \cite{Rex2007, Menzel2016}. Afterwards, we introduce the specific system which is investigated later on.

We consider a general system which consists of $N$ identical polar particles of mass $m$ with position and momentum coordinates $\Nr{}^N = \{\Nr{1}, \Nr{2},..., \Nr{N}\}$ and $\Np{}^N = \{\Np{1}, \Np{2},..., \Np{N}\}$. All particles self-propel with an active force$f_0$ into the direction of their orientation given by the unit vectors $\Nu{}^N = \{\Nu{1}, \Nu{2}, ..., \Nu{N} \}$. For convenience, we restrict the derivation to effectively two dimensional systems where an angle $\Nphi{i}$ determines the orientation $\Nu{i} = (\cos (\Nphi{i}), \sin (\Nphi{i}))^\top$. A particle's phase space coordinates are summarized as $\Nx{i} = \{ \Nr{i}, \Np{i}, \Nphi{i} \}$. The force $\mathbf{F}_i = \mathbf{F}^{(1)}(\Nr{i}) + \mathbf{F}^{(2)}(\Nr{}^N)$ summarizes an external potential force $\mathbf{F}^{(1)}(\Nr{i}) = -\nabla_{\Nr{i}} V^{\mathrm{ext}}(\Nr{i})$ and particle interactions $\mathbf{F}^{(2)}(\Nr{}^N) = -\sum_j \nabla_{\Nr{i}} V^{(2)}(\Nr{i}, \Nr{j})$ with the pair interaction potential $V^{(2)}(\Nr{i}, \Nr{j})$. Analogously to these translational forces particles change their orientation due to the torque $G_i(\Nr{}^N, \Nphi{}^N) = G^{(1)}(\Nr{i}, \Nphi{i}) + G^{(2)}(\Nr{}^N, \Nphi{}^N)$ with a one particle contribution $G^{(1)}(\Nr{i}, \Nphi{i})$ and a pairwise interaction torque $G^{(2)}(\Nr{}^N, \Nphi{}^N)$. The corresponding Langevin equations for the underdamped motion of the particle ensemble then read
\begin{eqnarray}
\eqalign{  \frac{\mathrm{d} \Nr{i}}{\mathrm{d} t} &= \frac{\Np{i}}{m}\\
\frac{\mathrm{d} \Np{i}}{\mathrm{d} t} &= - \gamma \, \Np{i} + \mathbf{F}_i(\mathbf{r}^N) + f_0 \, \Nu{i} + \sqrt{2D} \  \mathbf{\Gamma}_i(t)\\
\frac{\mathrm{d} \Nphi{i}}{\mathrm{d} t} &= G_i(\Nr{}^N, \Nphi{}^N) + \sqrt{2D_R} \ \xi_i(t)}
\end{eqnarray}
where $\gamma = \alpha / m$ is the damping constant with friction $\alpha$ and $D, D_R$ are translational and rotational diffusion constants. $\mathbf{\Gamma}_i(t), \, \xi_i(t)$ are independent $\delta$ - correlated white noises meaning $\langle \xi_i(t) \, \xi_j(t') \rangle = \delta_{ij} \, \delta(t-t')$ and analogously for the components of $\mathbf{\Gamma}_i$. Note that the damping $\gamma$ and diffusion $D$ are not necessarily related via the Stokes-Einstein relation since they might describe, e.g.,  the interaction with a substrate instead of a thermal bath.

These microscopic equations of motion for the stochastic state variables are converted into the corresponding Fokker-Planck equation of the $N$ body phase space probability density $f^{(N)}(\Nx{}^N, t)$ which gives us the probability of finding the system in a configuration $\Nx{}^N = \{ \Nr{}^N, \, \Np{}^N, \, \Nu{}^N \}$ at time $t$ \cite{risken1996fokker}. The resulting dynamical equation reads
\begin{eqnarray}
\eqalign{
\Npart{f^{(N)}} = \sum_{i=1}^N \Big[ - \frac{\Np{i}}{m} \cdot \dsub{\nabla}{\Nr{}}{i} f^{(N)} + \gamma \ \dsub{\nabla}{\Np{}}{i} \cdot \left(\Np{i} \ f^{(N)}\right) \\
\eqspacing - \mathbf{F}_i \cdot \dsub{\nabla}{\Np{}}{i} f^{(N)}  - f_0 \ \Nu{i} \cdot \dsub{\nabla}{\Np{}}{i} f^{(N)} \\
\eqspacing + D \ \dsub{\nabla}{\Np{}}{i}^2 f^{(N)} - \dsub{\partial}{\Nphi{}}{i} \left(G_i \ f^{(N)}\right) + D_R \ \dsub{\partial}{\Nphi{}}{i}^2 f^{(N)} \Big].}
\label{eq:model_N_body_fokker_planck}
\end{eqnarray}
This $N$ body problem is simplified by defining the $n$ body reduced phase space distribution functions

\begin{eqnarray}
\eqalign{
f^{(n)}(\Nx{}^n, t) =& \ \frac{N!}{(N-n)!} \Nint{}{}{\Nr{}^{(N-n)}} \Nint{}{}{\Np{}^{(N-n)}} \\
&\times \Nint{}{}{\Nphi{}^{(N-n)}} \ f^{(N)}(\Nx{}^N, t)
}
\end{eqnarray}
where $N-n$ of the $N$ bodies' state variables are integrated out. Equivalently, integrating the $N$ body Fokker-Planck equation over $N-1$ sets of particle variables yields the dynamical equation for the one body distribution
\begin{eqnarray}
\eqalign{
\Npart{f^{(1)}} = - \frac{\Np{1}}{m} \cdot \dsub{\nabla}{\Nr{}}{1} f^{(1)} + \gamma \ \dsub{\nabla}{\Np{}}{1} \cdot \left(\Np{1} \ f^{(1)}\right) \\
\eqspacing - \mathbf{F}^{\mathrm{ext}} \cdot \dsub{\nabla}{\Np{}}{1} f^{(1)} - f_0 \ \Nu{1} \cdot \dsub{\nabla}{\Np{}}{1} f^{(1)} + D \ \dsub{\nabla}{\Np{}}{1}^2 f^{(1)} \\
\eqspacing - \Nint{}{}{\Nr{2}} \Nint{}{}{\Np{2}} \Nint{}{}{\Nphi{2}} \ \mathbf{F}^{(2)} \cdot \dsub{\nabla}{\Np{}}{1} f^{(2)} \\
\eqspacing- \dsub{\partial}{\Nphi{}}{1} \left(G^{(1)} \ f^{(1)}\right) + D_R \ \dsub{\partial}{\Nphi{}}{1}^2 f^{(1)} \\
\eqspacing - \Nint{}{}{\Nr{2}} \Nint{}{}{\Np{2}} \Nint{}{}{\Nphi{2}} \ \dsub{\partial}{\Nphi{}}{1} \left(G^{(2)} \ f^{(2)} \right)
}
\label{eq:model_one_body_fokker_planck}
\end{eqnarray}
where we assume that the $N$ body density and its first derivatives decay to zero for all $\Nr{i}, \Np{i} \to \infty$ and are periodic in $\Nphi{i}$. The latter assumption is also used for the torques $G_1, G_2$. The reduced one body phase space density $f^{(1)}(\Nx{1}, t)$ gives the probability of finding a particle at the position $\Nr{1}$ with momentum $\Np{1}$ and orientation $\Nu{1}$ at time $t$. Note that due to particle interactions, \eref{eq:model_one_body_fokker_planck} still depends on the two body density $f^{(2)}(\Nx{1}, \Nx{2}, t)$.

For the next step we define the mean field quantities number density $\rho$, momentum current $\mathbf{j}$ and orientation current $\rho \, \mathbf{P}$ with the mean orientation field $\mathbf{P}$ as
\begin{eqnarray}
\eqalign{
\rho(\Nr{1}, t) &= \Nint{}{}{\Np{1}} \Nint{}{}{\Nphi{1}} \ f^{(1)}(\Nx{1}, t)\\
\mathbf{j}(\Nr{1}, t) &= \Nint{}{}{\Np{1}} \Nint{}{}{\Nphi{1}} \ \frac{\Np{1}}{m} \ f^{(1)}(\Nx{1}, t)\\
\rho \, \mathbf{P}(\Nr{1}, t) &= \Nint{}{}{\Np{1}} \Nint{}{}{\Nphi{1}} \ \Nu{1} \ f^{(1)}(\Nx{1}, t).}
\label{eq:model_mean_fields_definition}
\end{eqnarray}
Our goal is now to reduce the equation for $f^{(1)}$ to equations for the mean fields and simplify the problem further from there. \Eref{eq:model_one_body_fokker_planck} may be integrated over $\Np{1}$ and $\Nphi{1}$ which yields the continuity equation for the number density
\begin{equation}
\eqalign{
\Npart{\rho} + \dsub{\nabla}{\Nr{}}{1} \cdot \mathbf{j} = 0.}
\label{eq:model_continuity}
\end{equation}
Similarly, an equation for $\mathbf{j}$ is found by multiplying \eref{eq:model_one_body_fokker_planck} with $\Np{1} / m$ and then integrating over $\Np{1}$ and $\Nphi{1}$ to obtain
\begin{eqnarray}
\eqalign{
\Npart{\mathbf{j}} = &- \frac{1}{m^2} \Nint{}{}{\Np{1}} \Nint{}{}{\Nphi{1}} \ \Np{1} \left(\Np{1} \cdot \dsub{\nabla}{\Nr{}}{1} \right) f^{(1)} - \gamma \, \mathbf{j} \\
&+ \frac{1}{m} \, \rho \, \mathbf{F}^{\mathrm{ext}} + \frac{1}{m} \Nint{}{}{\Nr{2}} \ \mathbf{F}^{(2)} \, \rho^{(2)} + \frac{f_0}{m} \, \rho \, \mathbf{P}.}
\label{eq:model_current_dynamics}
\end{eqnarray}
Here, the two body number density
\begin{equation}
\eqalign{
\rho^{(2)}(\Nr{1}, \Nr{2}, t) = \Nint{}{}{\Np{1}} \Nint{}{}{\Np{2}} \Nint{}{}{\Nphi{1}} \Nint{}{}{\Nphi{2}} \ f^{(2)}}
\end{equation}
is used in the interaction integral.

Now, we effectively make the same two approximations as in \cite{Archer2009} to proceed further. The only difference is the dependence of $f^{(1)}$ on $\Nphi{1}$. First, we approximate the interaction integral containing the two body number density via
\begin{equation}
\eqalign{
\Nint{}{}{\Nr{2}} \ \mathbf{F}^{(2)}(\Nr{1}, \Nr{2}) \, \rho^{(2)} = - \rho \,  \dsub{\nabla}{\Nr{}}{1} \frac{\delta \Nfunctional_{exc} [\rho]}{\delta \rho}}
\label{eq:model_approx_functional}
\end{equation}
which is exact only in equilibrium. Therefore, the commonly used approximation is to use this expression also in the non-equilibrium case \cite{marconi1999dynamic}. $\Nfunctional_{exc}$ is the excess free energy which contains energy contributions due to particle interactions. The second approximation is to assume that the $\Np{1}$ and $\Nphi{1}$ dependencies in $f^{(1)}$ decouple and that the momentum part takes the 'local-equilibrium' Maxwell-Boltzmann form \cite{hansen1990theory}. The latter assumes that momentum is gaussian distributed around the local mean $\overline{\Np{1}} = m \, \mathbf{v}(\Nr{1},t)$ where $\mathbf{v}$ is the local average particle velocity analogous to the local mean orientation $\mathbf{P}$. Therefore, we write the one body density as
\begin{equation}
\eqalign{
f^{(1)}(\Nx{1}, t) = \frac{\Phi(\Nr{1}, \Nphi{1}, t)}{2\pi m kT} \ \exp\left( - \frac{\left(\Np{1} - m \, \mathbf{v}\right)^2}{2mkT} \right)}
\label{eq:model_approx_MB}
\end{equation} 
where $\Phi(\Nr{1}, \Nphi{1}, t)$ contains the orientational dependence and $kT$ is the thermal energy. With this form of the one body density it follows from the definition of the mean field densities that
\begin{eqnarray}
\eqalign{
\rho(\Nr{1}, t) &= \Nint{}{}{\Nphi{1}} \ \Phi(\Nr{1}, \Nphi{1}, t)\\
\mathbf{j}(\Nr{1}, t) &= \rho(\Nr{1}, t) \, \mathbf{v}(\Nr{1}, t).}
\label{eq:model_mean_fields_MB_approx}
\end{eqnarray}
Applying now the approximations \eref{eq:model_approx_functional} and \eref{eq:model_approx_MB} to the current equation in \eref{eq:model_current_dynamics} results in a dynamical equation for $\mathbf{v}$ which only depends on the other mean fields. This calculation is done and described in \cite{Archer2009} and therefore not rewritten here. All terms originating from the additional particle orientation vanish up to a coupling term in the momentum equation which does not change the principal calculation but is also found again in the resulting mean velocity equation
\begin{equation}
\eqalign{
\Npart{\mathbf{v}} + (\mathbf{v} \cdot \nabla) \mathbf{v} = -\gamma \, \mathbf{v} - \frac{1}{m} \nabla \frac{\delta \Nfunctional [\rho]}{\delta \rho} + \frac{f_0}{m} \, \mathbf{P}.}
\label{eq:model_mean_v_eq}
\end{equation}
Diffusion, interactions with other particles and external fields are summarized here under the Helmholtz free energy functional \cite{hansen1990theory}
\begin{equation}
\eqalign{
\Nfunctional [\rho] =& \ kT \Nint{}{}{\Nr{}} \ \rho \, \left[ \ln(\Lambda \, \rho) - 1\right] + \Nfunctional_{exc}[\rho] \\
&+ \Nint{}{}{\Nr{}} \ \rho \, V^{\mathrm{ext}}(\mathbf{r}).}
\end{equation}
The first term is the ideal gas free energy with the thermal wave length $\Lambda$.

Next, we find a dynamical equation for the mean orientation $\mathbf{P}$ by first multiplying equation \eref{eq:model_one_body_fokker_planck} with $\Nu{1}$ and then integrating over $\Np{1}$ and $\Nphi{1}$. The terms of the resulting dynamical equation of $\mathbf{P}$ are calculated separately in the following. First, we rewrite the left hand side of equation \eref{eq:model_one_body_fokker_planck} by using \eref{eq:model_continuity} and the expression for the current in equation \eref{eq:model_mean_fields_MB_approx} to obtain
\begin{equation}
\eqalign{
\Npart{(\rho \mathbf{P})} = \rho \, \Npart{\mathbf{P}} - \rho \, \mathbf{P} \, \nabla \cdot \mathbf{v} - \mathbf{P} \, (\mathbf{v} \cdot \nabla \rho).}
\end{equation}
Afterwards, we simplify the first term on the right hand side with the approximation \eref{eq:model_approx_MB} to
\begin{equation}
\eqalign{
\frac{1}{m} \Nint{}{}{\Np{1}} \Nint{}{}{\Nphi{1}} \ \Nu{1} \left(\Np{1} \cdot \dsub{\nabla}{\Nr{}}{1} \right) f^{(1)} \\
\eqspacing = \rho \, (\mathbf{v} \cdot \nabla) \mathbf{P} + \mathbf{P} (\mathbf{v} \cdot \nabla \rho) + \rho \, \mathbf{P} \, \nabla \cdot \mathbf{v}.}
\end{equation}
Like the pair interaction in the current dynamics \eref{eq:model_current_dynamics} we now approximate the interaction term for the orientation via an equilibrium excess functional 
\begin{eqnarray}
\eqalign{
\Nint{}{}{\Nr{2}} \Nint{}{}{\Np{2}} \Nint{}{}{\Nphi{2}} \ G^{(2)}(\Nr{1}, \Nr{2}, \Nphi{1}, \Nphi{2}) \ f^{(2)}\\
\eqspacing = - f^{(1)} \ \dsub{\partial}{\Nphi{}}{1} \frac{\delta \Nfunctional_{exc} [f^{(1)}]}{\delta f^{(1)}}}
\label{eq:model_approx3}
\end{eqnarray}
leading to the expression for the orientational interaction
\begin{eqnarray}
\eqalign{
-\Nint{}{}{\Nr{2}} \Nint{}{}{\Np{1}} \Nint{}{}{\Np{2}} \Nint{}{}{\Nphi{1}} \Nint{}{}{\Nphi{2}} \ \Nu{1} \, \dsub{\partial}{\Nphi{}}{1} \left( G_2 \, f^{(2)} \right)\\
\eqspacing = - \Nint{}{}{\Nphi{1}} \left( \dsub{\partial}{\Nphi{}}{1} \Nu{1} \right) \ \Phi \ \dsub{\partial}{\Nphi{}}{1} \frac{\delta \Nfunctional_{exc} [\Phi]}{\delta \Phi}\\
\eqspacing = - \rho \frac{\delta \Nfunctional_{P} [\mathbf{P}]}{\delta \mathbf{P}}.}
\end{eqnarray}
To arrive at the second line the momentum integrations on the two body density are carried out, partial integration on $\Nphi{1}$ is used and the approximation \eref{eq:model_approx3} is inserted. In the last step the excess functional $\Nfunctional_{\mathrm{exc}} [\Phi] = \Nfunctional_P[\mathbf{P}[\Phi]]$ is expressed via the mean orientation $\mathbf{P}$ by applying the chain rule for functional differentiation
\begin{equation}
\eqalign{
\frac{\delta \Nfunctional_{\mathrm{exc}} \left[\mathbf{P}[\Phi]\right]}{\delta \Phi(\Nr{}, \Nphi{}, t)} &= \Nint{}{}{\Nr{}'} \Nint{}{}{t'} \ \frac{\delta \Nfunctional_P [\mathbf{P}]}{\delta \mathbf{P}(\Nr{}', t')} \ \frac{\delta \mathbf{P}[\Phi](\Nr{}', t')}{\delta \Phi(\Nr{}, \Nphi{}, t)}\\
&= \frac{\delta \Nfunctional_P[\mathbf{P}]}{\delta \mathbf{P}(\Nr{},t)} \, \Nu{}(\Nphi{}).}
\end{equation}
Here, $\mathbf{P}[\Phi]$ is given by inserting the approximation \eref{eq:model_approx_MB} into the definition of $\mathbf{P}$ in equation \eref{eq:model_mean_fields_definition} which yields
\begin{equation}
\eqalign{
\mathbf{P}[\Phi] = \Nint{}{}{\Nphi{}} \ \Nu{} \, \Phi(\Nr{}, \Nphi{}, t).}
\end{equation}
All other terms in the resulting dynamical equation of $\mathbf{P}$ ever vanish or are dealt with via partial integration. A resulting non-vanishing term is the average one particle torque
\begin{eqnarray}
\overline{\mathbf{G}^{(1)}} = \frac{1}{\rho} \Nint{}{}{\Nphi{}} \, (\partial_{\Nphi{}} \Nu{}) \, G^{(1)} \, \Phi
\end{eqnarray}
which describes a particle's behaviour of orienting independently of other particles' orientations. Together with equations \eref{eq:model_continuity} and \eref{eq:model_mean_v_eq} we can now write down the final form of our model for underdamped active systems
\begin{eqnarray}
\eqalign{
\Npart{\rho} &= - \nabla \cdot (\rho  \textbf{v})\\
\frac{\mathrm{D} \textbf{v}}{\mathrm{D} t} &= -\gamma \, \textbf{v} - \frac{1}{m} \nabla \frac{\delta \Nfunctional}{\delta \rho} + \gamma v_0 \textbf{P}\\
\frac{\mathrm{D} \textbf{P}}{\mathrm{D} t} &= -D_R \textbf{P} - \frac{\delta \Nfunctional_P}{\delta \textbf{P}} + \overline{\textbf{G}^{(1)}}}
\label{eq:model_for_ud_active_systems}
\end{eqnarray}
with the convective derivative $\frac{\mathrm{D} }{\mathrm{D} t} = \frac{\partial}{\partial t} + (\textbf{v} \cdot \nabla)$. The self-propulsion velocity $v_0 = f_0 / \alpha$ corresponds to the steady state velocity of a particle in an environment with friction constant $\alpha$ and accelerated by the active force $f_0$. Although closure relations for particle interactions do not need to be formulated in a functional form this conceptual connection to equilibrium physics proofs to be a useful and instructive ansatz in the context of active matter models \cite{Rex2007, doostmohammadi2018active}.

\subsection{System}

By orienting on minimal ingredients for phenomenological models on the microscopic scale \cite{Dunkel2013, james2018turbulence, bratanov2015new, Ariel2018, Heidenreich2016, wensink2012meso} we now use equation \eref{eq:model_for_ud_active_systems} to specify a concrete system by choosing
\begin{eqnarray}
\eqalign{
\Nfunctional[\rho] &= \Nint{}{}{\Nr{}} \ \frac{c}{2} \, \rho^2\\
\Nfunctional_{P}[\mathbf{P}] &= \Nint{}{}{\Nr{}} \ \frac{\mathbf{P}}{2} \left[ -a + \lambda \left(q_0^2 + \nabla^2 \right)^2 \right] \mathbf{P} + \frac{\beta}{4} \, \left|\mathbf{P}\right|^4\\
\overline{\textbf{G}^{(1)}} &= 0.
}
\label{eq:functionals_choice}
\end{eqnarray}
Our model allows density variations which occur preferably in dilute systems. We consider repulsive interactions with an effective finite compressibility of the system described by $\Nfunctional$ where $c$ is the compressibility parameter. The interaction of orientations given by $\Nfunctional_{P}$ has the form of a Swift-Hohenberg functional \cite{Swift1977} with the preferred wave number for structure formation $q_0$ and additional free parameters $a, \lambda, \beta$. Its physical meaning becomes more evident when inserting equation \eref{eq:functionals_choice} into \eref{eq:model_for_ud_active_systems} resulting in 
\begin{eqnarray}
\eqalign{
\Npart{\rho} &= - \nabla \cdot \left(\rho \, \textbf{v}\right)\\
\frac{\mathrm{D} \textbf{v}}{\mathrm{D} t} &= \frac{1}{m} \left(- \alpha \, \textbf{v} - c \, \nabla \rho + \alpha v_0 \, \textbf{P} \right)\\
\frac{\mathrm{D} \textbf{P}}{\mathrm{D} t} &=  \left(\kappa - \beta \, |\mathbf{P}|^2\right) \, \mathbf{P} - \lambda \left(q_0^2 + \nabla^2 \right)^2 \ \mathbf{P}.
}
\label{eq:model_ud_orientation}
\end{eqnarray}
The $\kappa = a - D_R$ and $\beta$ terms were introduced before by Toner and Tu \cite{Toner1998} for the description of flocking in active systems. For $\kappa > 0$, the system evolves into the isotropic state $\mathbf{v} = \mathbf{0}$. We restrict to the non-trivial case $\kappa < 0$ where the local alignment strength of orientations $a$ outweighs rotational diffusion $D_R$, resulting in a preferred net velocity amplitude $\sqrt{\kappa / \beta}$. The second term in the orientation dynamics is interpreted as an anti-alignment interaction with strength $\lambda$ at a preferred particle distance $l_0 / 2 = \pi / q_0$ \cite{grossmann2014vortex}. This type of interaction supports the formation of dynamical patterns that are aligned at short distances, but usually do not favor a globally aligned state due to the preference of the anti-alignment on longer distances. Additional one body torques $G^{(1)}$ are not considered.

We are especially interested in the consequences of introducing inertia into active systems. Therefore, as an important reference case we determine the overdamped limit of \eref{eq:model_ud_orientation} by taking a low-mass limit. The convective derivative of the velocity equation can then be neglected leading to a quasi-stationary velocity dynamics. Additionally, we neglect convection of $\mathbf{P}$ which also results from the inclusion of inertia. With this, we simplify the dynamics in the overdamped limit to
\begin{eqnarray}
\eqalign{
\Npart{\rho} &= \frac{c}{2 \alpha} \, \nabla^2 \rho^2 - v_0 \, \nabla \cdot \left(\rho \, \textbf{P} \right)\\
\Npart{\mathbf{P}} &=  \left(\kappa - \beta \, |\mathbf{P}|^2 \right) \, \mathbf{P} - \lambda \left(q_0^2 + \nabla^2 \right)^2 \ \mathbf{P}.
}
\label{eq:model_od_orientation}
\end{eqnarray}
We rescale and numerically implement equations \eref{eq:model_ud_orientation} and \eref{eq:model_od_orientation} as described in \ref{sec:appendix_rescaling}. This reduces the set of dimensionless and independent parameters to $v_0, \lambda$ for the overdamped and to $m, v_0, \lambda$ for the underdamped model.

\section{Results}
\label{sec:results}

\begin{figure*}[t]
\centering
 \includegraphics[width=\textwidth]{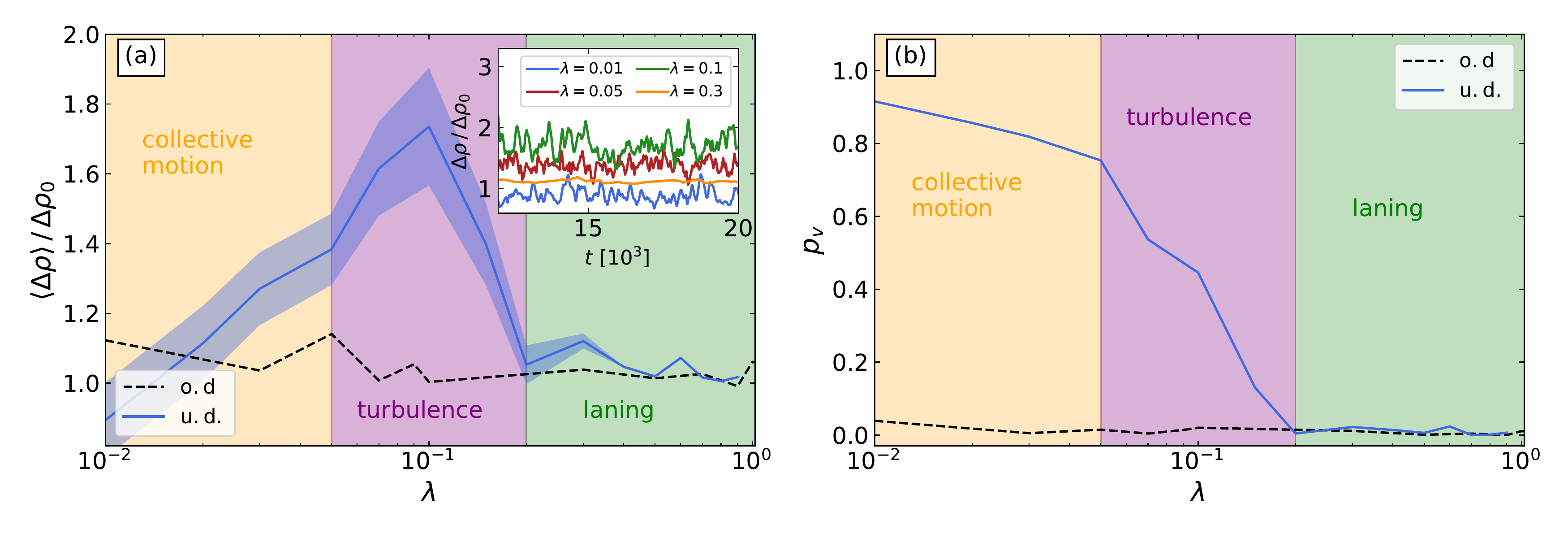}
\caption{Average density fluctuation $\langle \Delta \rho \rangle$ with its variance (a) and polar velocity order $p_v$ (b) distinguish qualitatively different states in the underdamped model (u.d.) indicated by background color: laning (green), turbulence (purple), and collective motion (orange). The inset in (a) shows exemplary courses of $\Delta \rho$ over simulation time. From the laning state to the turbulent transition regime the spatial fluctuation $\langle \Delta \rho \rangle$ and its temporal variance increase spontaneously at the critical value $\lambda = 0.2$ and a onset of global velocity ordering is observed in $p_v$. During the continuous transition to collective motion, $\langle \Delta \rho \rangle$ decreases again and global velocity order $p_v \approx 1$ is reached. Within the overdamped model (o.d.) neither an increase in spatio-temporal density fluctuations nor a onset of global velocity ordering is observed since the laning state, although increasingly distorted by vortical defects, persists over the hole tested $\lambda$ regime. Activity is set to $v_0 = 0.2$ and $m = 0.1$ in the underdamped case.}
\label{fig:observables}
\end{figure*}

We investigate the role of inertia by comparing predicted states along different regimes of the anti-alignment strength $\lambda$ in the under- as well as the overdamped model. The influence of the active drive $v_0$ and mass $m$ on the state diagram is discussed afterwards.

We distinguish qualitatively different states from changes of observables for density fluctuation and velocity alignment. First, we use the variance of the space averaged density
\begin{eqnarray}
\Delta \rho ^2 = \overline{\rho^2} - \overline{\rho}^2
\label{eq:space_variance}
\end{eqnarray}
and determine its time avarage
\begin{eqnarray}
\langle \Delta \rho \rangle = \frac{1}{T} \Nint{0}{T}{t} \ \Delta \rho(t)
\label{eq:density_fluctuations}
\end{eqnarray}
for a time interval $T$ that is chosen long enough such that the average does not depend on $T$. We use the mean value $\langle \Delta \rho \rangle$ to quantify density fluctuations in space while the corresponding variance is a measure for temporal fluctuations.

Second, we measure velocity alignment with the space averaged polar orientational order of the normalized velocity field
\begin{eqnarray}
p_v = \norm{ \overline{\mathbf{v} / \norm{\mathbf{v}} }}
\end{eqnarray}
where $\norm{\cdot}$ denotes the cartesian norm. A value near one indicates global orientational order of the velocity field while zero indicates the absence of global ordering. In the overdamped model the velocity field for $p_v$ is given from the density dynamics in equation \eref{eq:model_od_orientation} which has the form of a continuity equation. Thus, we have
\begin{eqnarray}
\mathbf{v} = - \frac{c}{2 \alpha} \, \nabla \rho + v_0 \, \mathbf{P}.
\label{eq:results_v_od}
\end{eqnarray}
We measure average density fluctuation $\langle \Delta \rho \rangle$ and velocity order $p_v$ for varying $\lambda$. From the shown results in figure~\ref{fig:observables} three states can be distinguished within the underdamped model which are discussed later in more detail. First, for high anti-alignment strengths the average density fluctuation is close to the reference value $\Delta \rho_0$ and the temporal variance is negligible. The system shows no global ordering of velocities. In this regime we observe alternating high and low density lanes along which particles move in opposite directions which we refer to as laning \cite{wensink2012emergent, mccandlish2012spontaneous, Menzel2013a, abkenar2013collective}. This is the only state which is equally predicted by the overdamped model. Second, near the critical value $\lambda = 0.2$, $\langle \Delta \rho \rangle$ and its temporal variance spontaneously increase which is accompanied by an onset of global orientational velocity ordering in $p_v$. We refer to this transition state as turbulent due to its non-steady character. And third, in the low $\lambda$ regime $\langle \Delta \rho \rangle$ continuously decreases again relative to the turbulent state and even below the value of the laning state. The temporal variance stays high. In this regime global ordering of velocities is observed with $p_v$ close to one leading to the identification of this state as collective motion. We now first discuss the laning state in the overdamped case in order to compare with the underdamped model afterwards.

\subsection{Laning in the overdamped model}
\label{sec:results_od}

\begin{figure}
  \centering
  \includegraphics[width=\columnwidth]{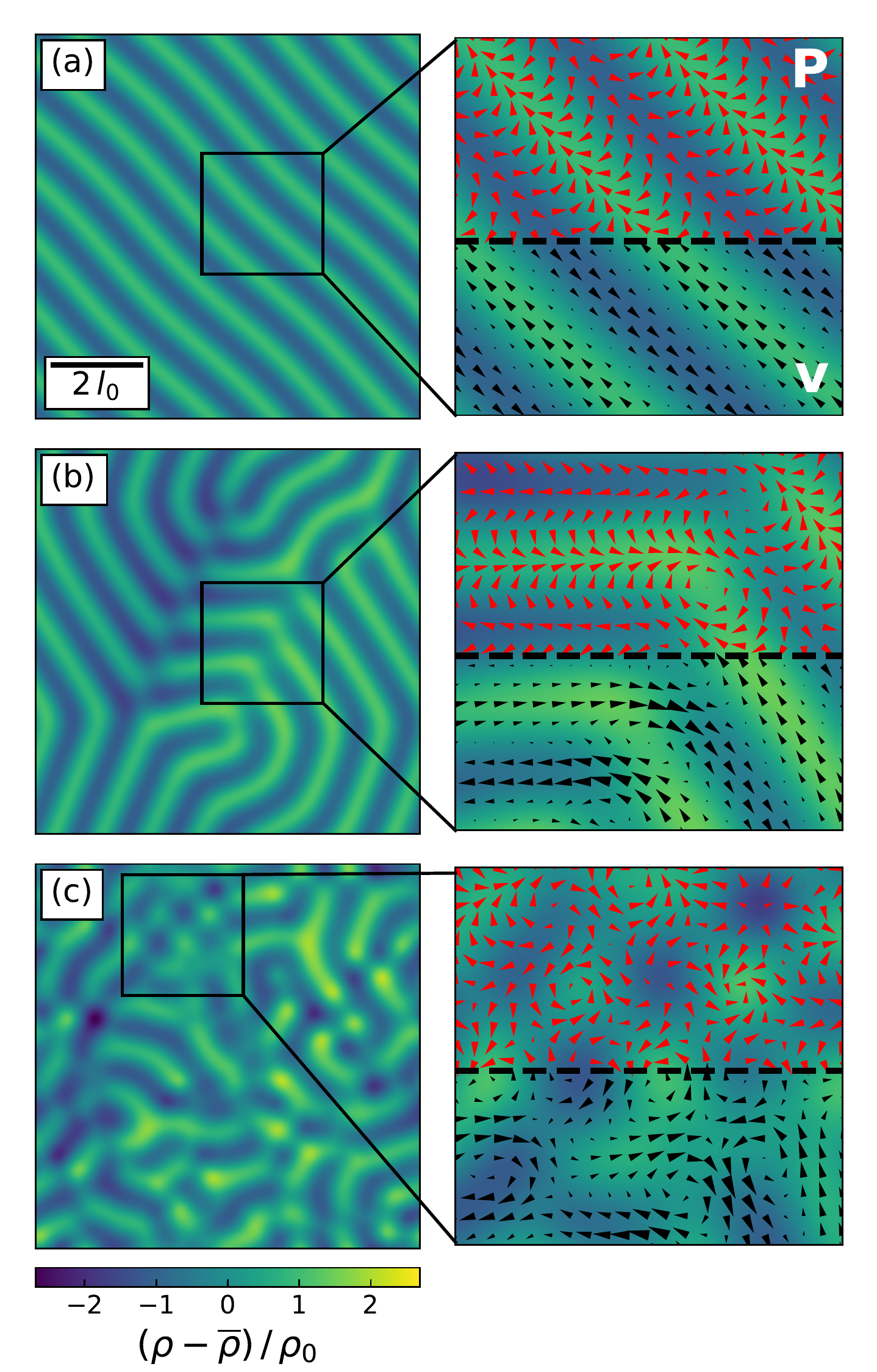}
  \caption{Density field predicted by the overdamped model for different anti-alignment strengths. The magnified regions show the orientation field (red) and the velocity field (black). (a) For high anti-alignment strengths the system is likely to reach a global laning state of alternating high and low density bands along which particles move within lanes. The instability in the density is driven by activity and balanced by the system's compressibility. Due to the preferred anti-alignment of orientations particles in neighbouring lanes self-propel in opposite directions with equal velocity causing a net particle current in the direction of movement of the high density lanes. (b) For intermediate $\lambda$ values we observe laning domains separated by metastable vortical defect boundaries in the orientation field. (c) In the low $\lambda$ regime the number of vortical defects in the orientation field increases further thereby preventing the formation of larger laning domains. Several vortivces might arange into a local square lattices where we observe circular motion. Parameters are set to (a) $\lambda = 3$, (b) $\lambda = 0.5$, (c) $\lambda = 0.03$, and $v_0 = 0.2$ for all systems.}
  \label{fig:od_states}
\end{figure}

The interaction of particle orientations given by $\Nfunctional_{P}$ in \eref{eq:functionals_choice} has two contributions. The Toner-Tu local alignment and the anti-alignment at distance $l_0 / 2$.  In figure~\ref{fig:od_states}~(a) we observe that for a high anti-alignment strength $\lambda$ the system favors a periodic modulation of orientations. The resulting orientation field accumulates density in bands with the same periodicity. This process is balanced by the compressibility of the system which counter acts along the arising density gradients. In the shown steady state the density and orientation fields may be approximated by harmonic modulations around their mean along one spatial direction. From inserting these expressions in the steady state force balance condition we predict the amplitude of the density variation and the corresponding spatial variance to
\begin{eqnarray}
\eqalign{
\rho_0 =& \sqrt{\frac{\kappa}{\beta}} \ \frac{v_0 \, \alpha}{q_0 \, c},\\
\Delta \rho_0 =& \frac{\rho_0}{\sqrt{2}}
}
\label{eq:results_od_rho0}
\end{eqnarray}
which is in accordance with our numerically found global stripe pattern and its spatial fluctiuation in figure~\ref{fig:observables}~(a). Along the maxima and minima of the density bands the active drive is not balanced by the systems compressibility and therefore induces particle fluxes organized in lanes. Due to the preferred anti-alignment particles in neighbouring high and low density lanes propel in opposite directions with the same self-propulsion velocity $v_0$ which explains the observed absence of global velocity ordering in figure~\ref{fig:observables}~(b). However, we emphasize that the density difference between the opposite propulsion directions leads to a net current in the direction of movement within high density lanes. For high enough active drive we expect all particles might accumulate in the high density lanes and move in the same direction, thereby maximizing the current, which is however not explicitly tested. Such unidirectional laning states are observed in particle simulations with orientational alignment \cite{Menzel2013a}. There, the maximum distance of local alignment coincides with the resulting lane distance. In this reference the formation of lanes is explained as an overreaction of the alignment interaction. This differs from the laning mechanism observed here since alignment of particles happens only locally and an anti-alignment rule dominates at further distances thereby determining the lane spacing.

When the anti-alignment strength $\lambda$ is lowered the orientation field becomes more likely to locally form vortices with diameter $l_0 / 2$. Seeing the orientational interactions as the derivative of the vectorial PFC functional $\Nfunctional_P$ the increasing occourence of vortices between laning domains can be seen as a transition from the stripe to the crystal phase. The difference to a typically used one component PFC interaction \cite{elder2002modeling, elder2004modeling, emmerich2012phase} is the coupling of the two vector components of $\mathbf{P}$. So instead of a clear transition from laning (stripe phase) to a regular lattice of vortices (crystal phase) we observe in figure~\ref{fig:od_states}~(b) that the system is typically stuck in metastable states of laning domains which are separated by defects in the form of vortices in the orientation field. Those defects increase the functional free energy of the $\lambda$ term relative to the global laning state as anti-parallel alignment is only locally given.

For the lowest tested $\lambda$ values the number of vortices in the orientation field steadily increases until clear laning domains are absent. Instead the density pattern varies locally depending on the metastable state of the orientation field. Additionally, we find several vortices in the orientation field which arrange in local lattices of alternating clockwise and anti-clockwise motion like illustrated in figure~\ref{fig:od_states}~(c). Different to the hexagonal lattice symmetry arising for one component PFC functionals we solely observe square lattices which enable a frustration free arrangement of the vortices' orientation of circular motion. We finally note that local square lattices are typically distorted by their surrounding leading to inward and outward spiraling vortices in the orientation field instead of perfectly circular ones. Inward spiraling orientations accumulate density while outward spiraling ones spread density. In the steady state the resulting density variations between contrary rotating vortices are balanced by the compressibility of the system resulting in square lattices of perfectly circular vortices in the velocity field.

\subsection{States in the underdamped model}
\label{sec:results_ud}

\begin{figure}[t]
  \centering
  \includegraphics[width=\columnwidth]{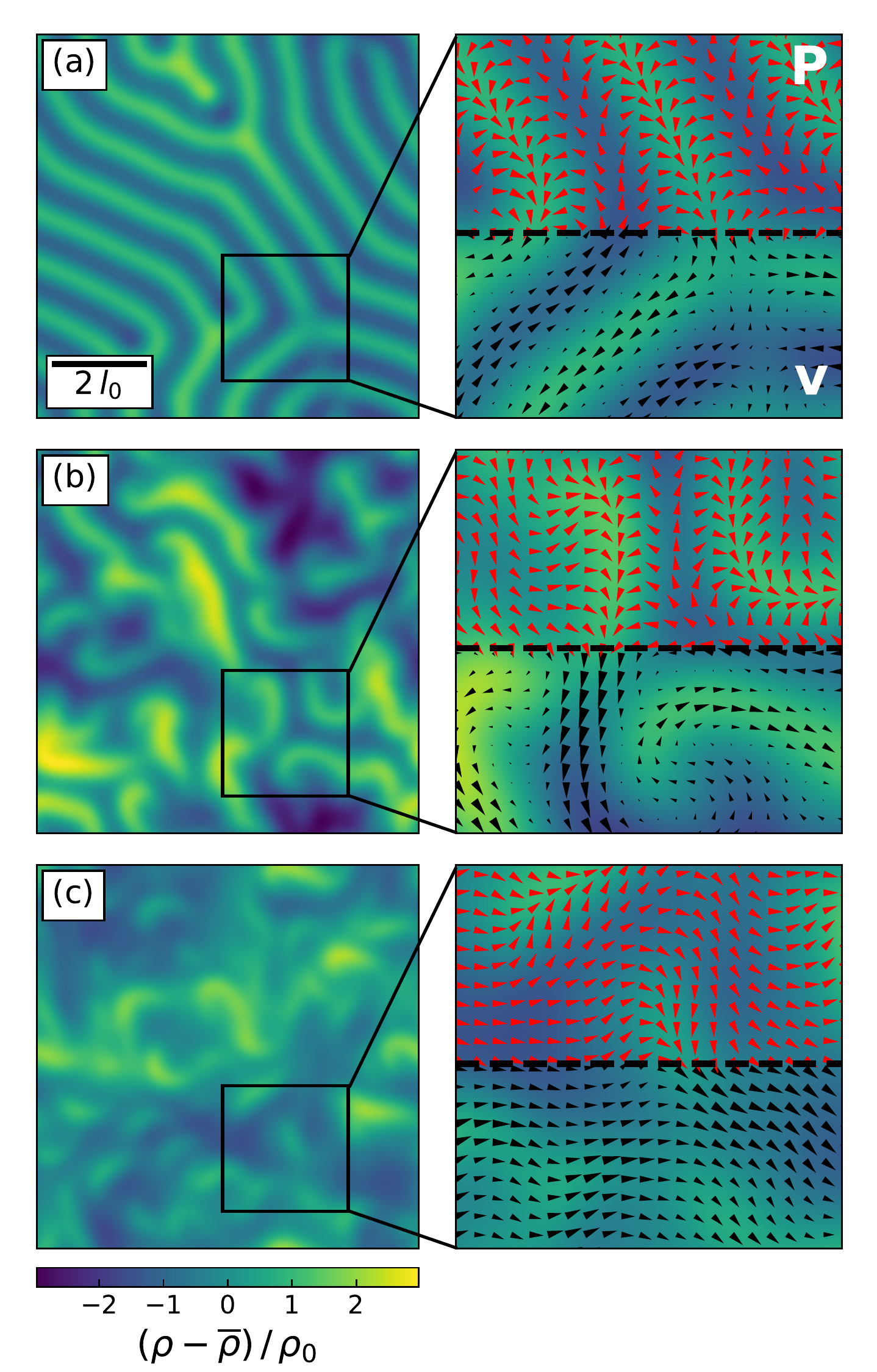}
  \caption{Density field predicted by the underdamped model for different anti-alignment strengths. The magnified regions show the orientation field (red) and the velocity field (black). (a) For high anti-alignment strengths the same laning state as in the overdamped model is found. However, for intermediate values vortex defects in the orientation field become unstable due to convection. Instead, differently orientated laning domains connect through linkage and branching of lanes. (b) Below the critical value $\lambda = 0.2$ convective flows also destabilize laning structures. In this turbulent transition state the interplay of lane formation and convection results in strong spatio-temporal fluctuations of the density. (c) For the lowest observed $\lambda$ local alignment of orientations dominates and a global state of collective motion arises slightly modulated by the weak anti-alignment and resulting density fluctuations. System parameters are set to (a) $\lambda = 0.6$, (b) $\lambda = 0.1$, (c) $\lambda = 0.02$, and $(m, v_0) = (0.1, 0.2)$ for all systems.}
\label{fig:ud_states}
\end{figure}

\paragraph*{Laning.} For high $\lambda$ values the laning state found within the overdamped model is equally predicted in the underdamped model. We find the same value of spatial fluctuation $\Delta \rho_0$ and absence of global velocity ordering $p_v = 0$ (see figure~\ref{fig:observables}). We find vortical defects in the orientation field for anti-alignment strengths $\lambda \gtrsim 1$. However, in contrast to the overdamped limit they are absent when $\lambda$ is lowered as exemplary shown in figure~\ref{fig:ud_states}~(a). We explain this difference with the convection of orientation. A vortex in the orientation field induces a corresponding circulating flux. Due to their inertia particles radially leave the vortex which gets distorted due to this convection. The anti-alignment interaction counteracts this distortion similar to a centripetal force which holds objects on circular orbits. Consequently, if $\lambda$ is lowered enough the formation of vortices in the orientation field is inhibited. In the steady state the grain boundaries between different laning domains then resolve by branching and linkage of lanes with equivalent orientation at the boundary. In this sense inertial convection is actually beneficial for the formation of a global laning state at intermediate anti-alignment strengths since it heals out vortical defects in the orientation field.

\paragraph*{Turbulence.} Lowering $\lambda$ into the turbulent regime suddenly changes the dynamics considerably since now convective flows not only destabilize vortical defects in the orientation field but also the laning structures due to the weakened anti-alignment. As can be seen from the orientation field in figure~\ref{fig:ud_states}~(b) the laning structure evolves only locally. The system does no longer reach a steady state but instead large spatio-temporal fluctuations are self-sustained over time which is reflected by an increase in density fluctuation $\langle \Delta \rho \rangle$ and its temporal variance. The movie in the supplemental materials gives an impression of the fluctuating density field. The non-steady character of the turbulent state results from the continues interplay of local lane formation and their destabilization due to convective flows. The sudden increase of density fluctuations coincides with the onset of global velocity ordering in figure~\ref{fig:observables}. Since convective flows destabilize lane formation local alignment is no longer restricted to single lanes resulting in a small but global drift velocity which is reflected in a moderate velocity order $p_v$. Our finding that inertial convection destabilizes the laning state is verified by switching off the convective terms in the dynamical equations of velocity and orientation within the underdamped model. Then, we again find local laning structures and vortical defects like in the overdamped model where otherwise turbulence would arise.

\paragraph*{Collective Motion.} In the turbulent transition state anti-alignment is just weak enough so that laning is unstable but still strong enough to inhibit global velocity ordering. This changes when $\lambda$ is further lowered since then local alignment of orientations increasingly dominates over the weak anti-alignment resulting in the emergence of global orientational velocity order as measured in figure~\ref{fig:observables}~(b) and shown in figure~\ref{fig:ud_states}~(c). The homogeneous flow field transports particle density without accumulating it too much leading to small density fluctuations in figure~\ref{fig:observables}~(a) comparable to the laning state or even smaller for low enough $\lambda$. Since anti-alignment continually perturbs the homogeneous velocity field the temporal variance of $\langle \Delta \rho \rangle$ stays on a high level. We suggest that convective flows stabilize the global collective motion as they do not influence a homogeneous state of aligned orientations but mix any arising misaligned clusters with their surrounding due to convective transport of orientation. Hence, the formation of a larger misaligned cluster is suppressed by convection.

\subsection{Remarks}

We emphasize here that the convective flows necessary to destabilize the laning state in the underdamped model are driven by activity. Therefore, when lowering the active drive $v_0$ within the turbulent $\lambda$ regime shown in figure~\ref{fig:observables} we observe a continuous decrease of $\langle \Delta \rho \rangle$, its temporal variance, and $p_v$ back to the values characteristic of the laning state. Then, the known global lane structure as in figure~\ref{fig:ud_states}~(a) emerges again. However, due to the relatively weak anti-alignment convective flows still perturb the global structure causing it to continually rearrange over time.

Furthermore, we note that the value of the particle mass $m$ as a third free parameter is not relevant for the state diagram of the underdamped model. Especially for the low mass value $m=0.1$ we do not find the states of the overdamped model as a limit case but the qualitatively distinct ones discussed above. Also when choosing the mass two orders of magnitude higher, we find the same qualitative states as in figure~\ref{fig:ud_states} and the same values for the observables in figure~\ref{fig:observables}. We therefore conclude that the mass parameter is solely relevant for the intermediate dynamics, at least on the damping time scale $\gamma^{-1} = m / \alpha$.

\section{Conclusion}
\label{sec:conclusion}

In this work we have derived an effective continuum model for underdamped active matter based on a dynamical density functional theory for passive systems \cite{Archer2009}. Further, we applied the model to a system with local alignment and distant anti-alignment interaction of orientations. Instead of a continuous transition from a underdamped state diagramm to a overdamped one for $m \to 0$ the underdamped model predicts different states irrespective of the particle mass. Our numerical findings show that activity driven convective flows explain this qualitative difference.

The dynamical equations of our overdamped model \eref{eq:model_od_orientation} and \eref{eq:results_v_od} are similar to the those in \cite{grossmann2014vortex}. In both cases strong anti-alignment leads to structure formation, here in the form of laning and in \cite{grossmann2014vortex} as vortex array. This difference is due do the respective implementation of repulsion. In \cite{grossmann2014vortex} particles actively rotate their orientation to avoid high density regions while in the present model this repulsion, modeled by the compressibility, is a passive one in the sense that it lowers a particle's velocity if it moves to higher densities instead of turning its orientation. More importantly, for lower anti-alignment strength a transition to collective motion with a turbulent transition regime is observed in \cite{grossmann2014vortex} reminiscent to the findings within our underdamped model although their model is fully overdamped. The authors explain this with mesoscopic convective flows arising from the local alignment. We emphasize here that such effective convective effects in fully overdamped active systems necessarily arise from interactions or noise while in the present case their physical origin is inertia. Also the phenomenom of active turbulence in overdamped systems typically is driven from hydrodynamic interactions which are included in the dynamical equations with a term structurally equivalent to convection \cite{wensink2012meso, bratanov2015new, James2018}.

\appendix

\section{Rescaling and Implementation}
\label{sec:appendix_rescaling}
We rescale equations \eref{eq:model_ud_orientation} and \eref{eq:model_od_orientation} in order to extract physically relevant parameters and for numerical implementation. Independent quantities are rescaled to their dimensionless form according to the rules
\begin{eqnarray}
\eqalign{
\rescale{\Nr{}}{q_0},  &\rescale{t}{\kappa^{-1}},\\
\rescale{\rho}{\frac{\alpha \kappa}{c q_0^2}}, &\rescale{\mathbf{v}}{\frac{\kappa}{q_0}}, \\
\rescale{m}{\frac{\alpha}{\kappa}}, &\rescale{v_0}{\frac{(\kappa \beta)^{1/2}}{q_0}}, \\
\rescale{\mathbf{P}}{\left(\frac{\kappa}{\beta}\right)^{1/2}}, \qquad & \rescale{\lambda}{\frac{\kappa}{q_0^4}}
}
\label{eq:model:orientation_rescaling}
\end{eqnarray}
which leads to the dimensionless form of the underdamped model in equation \eref{eq:model_ud_orientation}
\begin{eqnarray}
\eqalign{
\Npart{\rho} &= - \nabla \cdot \left(\rho \, \textbf{v}\right)\\
\frac{\mathrm{D} \textbf{v}}{\mathrm{D} t} &= \frac{1}{m} \left( - \textbf{v} - \nabla \rho + v_0 \, \textbf{P} \right)\\
\frac{\mathrm{D} \textbf{P}}{\mathrm{D} t} &=  (1 - |\mathbf{P}|^2) \, \mathbf{P} - \lambda \left(1 + \nabla^2 \right)^2 \ \mathbf{P}.
}
\end{eqnarray}
And for the overdamped model in equation \eref{eq:model_od_orientation} we have
\begin{eqnarray}
\eqalign{
\Npart{\rho} &= \frac{1}{2} \nabla^2 \rho^2 - v_0 \, \nabla \cdot \left(\rho \, \textbf{P} \right)\\
\Npart{\mathbf{P}} &=  \left(1 - |\mathbf{P}|^2 \right) \mathbf{P} - \lambda \left(1 + \nabla^2 \right)^2 \ \mathbf{P}.
}
\end{eqnarray}
For numerical implementation we discretize all fields on a rectengular grid. The simulation box has lengths of multiples of $l_0$ and employs periodic boundary conditions. A pseudo-spectral algorithm is used for numerical time iteration. For the evaluation of the convective terms in real space a third order upwind scheme is applied \cite{swanson1992central}. Time stepping is implemented via a semi-implicit Euler discretization with fixed step size. All simulations are started from homogeneous initial conditions. The average density is always set to $\overline{\rho} = 1$.

\section*{References}

\bibliographystyle{iopart-num}
\bibliography{references}

\end{document}